\documentclass[a4paper,11pt]{article}
\usepackage{pos}
\renewcommand{\logo}{\relax}

\def \beq  {\begin{equation}}
\def \eeq  {\end{equation}}
\def \beqar {\begin{eqnarray}}
\def \eeqar {\end{eqnarray}}

\def\la {{\langle}}
\def\ra {{\rangle}}

\def\Tr {{\rm Tr}}
\def \tr {{\rm tr}}

\def\del {\partial}

\def\g {{\gamma}}
\def\d {{\delta}}

\def\d {\delta}

\def\l {\lambda}

\def\bz {{\bar{z}}}

\def\A {{\cal A}}

\def\D {{\cal D}}

\def\half{\textstyle{1\over 2}}

\mathchardef\mhyphen="2D

\def \CMP {{Commun. Math. Phys.}}
\def \PRL {{Phys. Rev. Lett.}}

\def \PR {{Phys. Rev.}}

\title{Aspects of higher dimensional quantum Hall effect: Bosonization, entanglement entropy}

\author*[a,b]{Dimitra Karabali}

\affiliation[a]{Physics and Astronomy Department, Lehman College, CUNY,\\
  Bronx, NY 10468, USA}

\affiliation[b]{The Graduate Center, CUNY,\\
New York, New York 10016, USA}

\emailAdd{dimitra.karabali@lehman.cuny.edu}

\abstract{
I give a brief review of higher dimensional quantum Hall effect (QHE) and how one can use a general framework to describe the lowest Landau level dynamics as a noncommutative field theory whose semiclassical limit leads to anomaly free bulk-edge effective actions in any dimension.
I then present the case of QHE on complex projective spaces and focus on the entanglement entropy for integer QHE in even spatial dimensions. In the case of $\nu=1$, a semiclassical analysis shows that the entanglement entropy is proportional to the phase-space area of the entangling surface with a universal overall constant, same for any dimension as well as abelian or nonabelian background magnetic fields. This is modified for higher Landau levels.}

\FullConference{%
  Corfu Summer Institute 2021 "School and Workshops on Elementary Particle Physics and Gravity"\\
  29 August - 9 October 2021\\
  Corfu, Greece
}

\begin{document}
\maketitle

\section{Introduction}

Quantum Hall effect (QHE) has long been a fascinating phenomenon.
It is associated with systems of nonrelativistic electrons in two spatial dimensions in the presence of a strong perpendicular magnetic field and low temperatures \cite{girvin}. The hallmark of QHE is the remarkable quantization of the Hall electrical conductivity $\sigma_H= \nu e^2 / h$, where the filling fraction $\nu$ takes integer or special fractional values.

In addition to a variety of important experimental results 
QHE has provided a theoretical framework
 to study the interplay of many interesting ideas such as noncommutative geometries and fuzzy spaces, bosonization, topological field theories, bulk-edge dynamics, etc. It turns out that these connections as well as most of the important features of the original two-dimensional QHE extend to higher dimensions and the results are quite interesting.
 
One of the first attempts to extend the QHE to four dimensions was by Hu and Zhang, who formulated QHE on $S^4$ \cite{HZ}. Right after that, in collaboration with Nair we extended this to arbitrary even dimensions by formulating QHE on the complex projective spaces  ${\mathbb{CP}}^k$ \cite{QH1}-\cite{karabali}. (For $k=1$, this reduces to the well known case of QHE on $S^2 \sim {\mathbb{CP}}^1$ \cite{Haldane}.) This extension introduces two important new features. First, higher dimensionality and second, the possibility of having both abelian and nonabelian background magnetic fields when $k > 1$. The nonabelian case is relevant for a many-body system of fermions with internal degrees of freedom. Subsequently, different versions of higher dimensional QHE have been developed by other groups in both even and odd spatial dimensional spaces \cite{others}. 
Over the last few years four-dimensional QHE has been experimentally realized in optical and cold atom systems, using the concept of synthetic dimensions \cite{exp}, providing an exciting interface between theory and experiment for higher dimensional QHE ideas. 

In this talk I will briefly outline how the dynamics of integer QHE in any dimension is described by a noncommutative bosonic field theory and how this has been used to construct effective actions that capture the response of the system to electromagnetic and/or gravitational fluctuations \cite{QH1}-\cite{QHEaction}. 
I will then briefly describe the spectrum of the QHE on ${\mathbb{CP}}^k$ and focus on the more recent work regarding the entanglement entropy for integer QHE in higher dimensions \cite{karabali}. 

\section{ Lowest Landau level dynamics as a noncommutative bosonic field theory}

When QHE is defined on a compact space, each Landau level provides a finite dimensional Hilbert space.
In general, the spectrum decomposes into distinct, degenerate Landau levels separated by an energy gap proportional to the external magnetic field. In the presence of a strong magnetic field the dynamics is restricted in the lowest Landau level (LLL). Let us assume that the dimension of the LLL is $N$ and $K$ of these states are occupied by fermions. For simplicity, we can ignore the spin degree of freedom so that each state can be occupied by a single fermion. The observables of the full theory projected onto the lowest Landau level are $(N \times N)$ matrices and the dynamics of all possible excitations within the LLL is then described by a one-dimensional matrix action. In a physical sample there is also a potential $\hat{V}$ which confines the fermions within the sample. The fermions are localized around the minimum of the potential but, because of the exclusion principle, they spread out and form droplets of almost constant density, the so-called quantum Hall droplets. The low energy excitations of these droplets, corresponding to transitions of outer electrons to states of higher energy within the same Landau level are boundary fluctuations, the so-called edge excitations. The ground state droplet is characterized by a diagonal density matrix $\hat {\rho}_0$, with diagonal elements equal to 1 for occupied states and 0 for unoccupied ones. The most general fluctuations whithin the LLL which preserve the number of occupied states are unitary transformations $\hat {\rho}_0 \rightarrow \hat{\rho}=\hat{U}  \hat {\rho}_0 \hat{U} ^ \dagger$, where  $\hat{U}$ is an $(N \times N)$ unitary matrix. $\hat{U}$ can be thought of as a ``collective" variable describing excitations within the LLL. The
action for ${\hat U}$ is given by 
\beq
S_{0}=  \int dt~ \Tr \left[ i  {\hat \rho}_0 { \hat U}^\dagger \del_t {\hat U}
~-~ {\hat \rho}_0 {\hat U}^\dagger {\hat{V}} {\hat U} \right]
\label{1}
\eeq
The equation of motion resulting from (\ref{1}) is the expected evolution equation for $\hat{\rho}$, namely
\beq
i {{\partial \hat{\rho}} \over {\partial t}} = [ \hat{V} , \hat{\rho}]
\label{3}
\eeq
It is worth emphasizing that $S_{0}$ is a universal matrix action with no explicit dependence on properties of space on which QHE is defined, the abelian or nonabelian nature of fermions, etc. One can bring in the space-time features by writing $S_0$ as a noncommutative field theory action, namely
\beqar
S_{0}&=&  \int dt~ \Tr \left[ i  {\hat \rho}_0 { \hat U}^\dagger \del_t {\hat U}
~-~ {\hat \rho}_0 {\hat U}^\dagger {\hat{V}} {\hat U} \right]  \label{matrix} \\
&= & {N \over N'} \int d\mu ~dt~ tr \left[ i ({\rho}_0 *{  U}^\dagger  * \del_t { U})
~-~ ({ \rho}_0 *{U}^\dagger  * {{V}} * {U}) \right]   \label{NC}
\eeqar
where $d \mu$ is the volume measure of the space where QHE has been defined and $\rho_0,~U,~V$ are the symbols of the corresponding matrices on this space. Equation (\ref{NC}) is written for the case of nonabelian fermions coupled to a background gauge field in some representation $J'$ of dimension $N'$; the corresponding symbols are $(N' \times N')$ matrix valued functions. 
Our notation is such that ``$\Tr$" indicates trace over the $N$-dimensional LLL Hilbert space while ``$\tr$" indicates trace over the $N'$-dimensional representation $J'$. In the case of abelian fermions, $N'=1$ and $\tr$ is trivial. 
In general the symbol and star-product are given by 
\beqar
O(\vec{x}, t) & = & { 1 \over N} \sum_{m,l} \Psi_m(\vec{x}) O_{ml}(t) \Psi^*_l(\vec{x})
\label{symb} \\
\big(\hat{O}_1\hat{O}_2\big)_{symbol} & = & O_1(\vec{x}, t) * O_2 (\vec{x}, t) 
\label{star}
\eeqar
where $\Psi_m(\vec{x})$, $m=1,\cdots,N$,  represent the correctly normalized LLL single-particle wavefunctions. 

This bosonization technique can be extended to include fluctuating gauge fields (in addition to the uniform background magnetic field which defines the Landau problem) by gauging the matrix action (\ref{1}), namely
\beqar
S &=& \int dt~ \Tr \left[ i  \hat {\rho}_0  \hat {U}^\dagger ( \del_t + i \hat{\A} ) \hat {U}
~-~  \hat {\rho}_0 \hat {U}^\dagger \hat{V} \hat{U} \right] \\
&=& {N \over N'} \int dt~d\mu ~ \tr~\left[ i  \rho_0 * U^\dagger  * \del_t U
~-~ \rho_0 *U^\dagger * (V + \A) *U \right]
\label{6}
\eeqar
An obvious question then is how $\A$ related to the gauge fields coupled to the original fermions? 
The action (\ref{6}) is invariant under the infinitesimal transformations
\beqar
\delta U & =& - i \l * U \nonumber \\
\delta \A (\vec{x}, t)  & = & \del _t \l (\vec{x}, t) - i \left( \l* (V+\A) - (V + \A) * \l \right) 
\label{11}
\eeqar
Since the action $S$ is supposed to describe gauge interactions of the original system it also has to be invariant under the usual gauge transformation
 \beqar
 \delta A_{\mu} & = & \del_{\mu} \Lambda + i [\bar{A}_\mu + A_\mu , ~ \Lambda] 
 \label{13} \\
 \delta \bar{A}_{\mu} & = & 0 \nonumber 
 \eeqar
 where $\Lambda$ is the infinitesimal gauge parameter and $\bar{A}_\mu$ is a possible uniform non-Abelian background field. For abelian gauge fields the commutator term in (\ref{13}) is zero. The fact that $S$ is invariant under (\ref{11}) and (\ref{13}) implies that the transformation (\ref{11}) can be thought of as a nonlinear realization of the gauge transformation (\ref{13}). This provides a clear way to construct (up to gauge invariant terms) $\A$ as a function of $A_{\mu}$ and derive the bosonized action of the LLL fermionic system in the presence of gauge interactions. Further, since ${\A}$ can be thought of as the time-component of a noncommutative gauge field, the relation between ${\A}$ and the commutative gauge fields $A_\mu$ is essentially a Seiberg-Witten transformation \cite{QH5,QH6}.
 
This approach provides a very general way to construct the bosonic action describing the dynamics of  the underlying LLL fermionic system in any space that admits a consistent formulation of QHE. At the semiclassical limit, where $N, K$ are large and $N \gg K$, the action (\ref{6}) splits into a chiral edge action that has support on the boundary of the droplet and a purely $A_\mu$-dependent Chern-Simons type bulk action.  Furthermore, although each of these actions is not gauge invariant there is an exact anomaly cancellation between the two, since the original action (\ref{6}) is by construction gauge invariant.  This is a generalization of a method used by Sakita \cite{sakita2} to derive the electromagnetic interactions of LLL spinless electrons in the two-dimensional plane. 
We have implemented this approach in a series of papers to construct effective actions for the higher dimensional integer QHE on ${\mathbb{CP}}^k$, whose single-particle spectrum is described in the next section. 
We found that, in the absence of fluctuating gauge fields, the edge dynamics of the higher dimensional chiral droplets is described by a higher dimensional generalization of the Wess-Zumino-Witten action \cite{QH2,QH3}. In the presence of fluctuating gauge fields the effective action splits into a bulk Chern-Simons type action and a gauged, chiral WZW action in a way that the full action is gauge invariant \cite{QH5,QH6}.

 We further extended the derivation of effective actions for higher dimensional QHE systems by including gravitational metric perturbations around background fields of a specified topological class, thus moving away from the standard metric of ${\mathbb{CP}}^k$ spaces \cite{QHEaction}. The analysis was based on the use of index theorems applicable for complex manifolds, in particular the Dolbeault index which essentially measures the degeneracy of the lowest Landau level and therefore the charge of the system if one assigns a unit charge to each nonrelativistic electron. Integrating the Dolbeault index density with respect to the gauge field, with some input from the standard descent procedure for anomalies, we were able to derive the bulk topological effective action for integer quantum Hall systems including gauge and gravitational interactions on manifolds of arbitrary even dimension. This extends previous two-dimensional results on effective actions and related transport coefficients \cite{otherQHE1,otherQHE2} to all even spatial dimensions.

\section { Quantum Hall effect on ${\mathbb {CP}}^k$}

In this section I will briefly describe the spectrum of QHE on ${\mathbb {CP}}^k$, following the group theoretic analysis developed in \cite{QH1}-\cite{QH3}. ${\mathbb {CP}}^k$ are $2k$-dimensional manifolds that can be locally parametrized by $k$ complex coordinates $z_i$. They can also be thought as coset manifolds, namely
\beq
{\mathbb{CP}^k} = {SU(k+1) \over U(k)}  
\label{coset}
\eeq
It turns out that a group theoretic approach provides a generic and straightforward way to extract the single particle spectrum for each Landau level. It is immediately clear from (\ref{coset}) that since $U(k) \sim U(1) \times  SU(k)$, one can have both $U(1)$ and $SU(k)$ background magnetic fields and further the Landau wavefunctions can be obtained as functions of $SU(k+1)$ with specific transformation properties under $U(k)$. A basis for such functions is given by the so-called Wigner $\cal{D}$-functions, which are the matrices corresponding to the group elements in the unitary irreducible representations, i.e.,
\beq
\Psi^{J} \sim \D^{J}_{L,R}(g) = \la J , l_A \vert\, g\,\vert J, r_A \ra \label {wigner}
\eeq
where $J$ denotes the representation
and $l_A, ~r_A$ stand for two sets of quantum numbers specifying the 
states within the representation. On an element $g\in SU(k+1)$, 
we can define left and right $SU(k+1)$ actions by
\beq
{\hat{L}}_A ~g = T_A ~g, \hskip 1in {\hat{R}}_A~ g = g~T_A
\eeq
where $T_A$ are the $SU(k+1)$ generators in the representation to which $g$ belongs. The left transformations ${\hat{L}}_A$ correspond to magnetic translations. There are $2k$ right generators of $SU(k+1)$ which are not in
$U(k)$; these can be separated into $T_{+i}$, $i=1,2 \cdots ,k$, which are
of the raising type and $T_{-i}$ which are of the lowering
type. These generate translations on ${\mathbb {CP}}^k$, while the right $U(k)$ transformations generate rotations at a point and correspond to gauge transformations. The covariant derivatives on ${\mathbb {CP}}^k$ are given by 
\beq
\D_{\pm i}  = i\,{{\hat R}_{\pm i} \over r}
\label{covariant}
\eeq
where $r$ can be thought of as the radius of ${\mathbb {CP}}^k$. 
This is consistent with the fact that the commutator of
covariant derivatives is the magnetic field.
The commutators of ${\hat R}_{+i}$ and ${\hat R}_{-i}$ are in the
Lie algebra of $U(k)$;  in the case of ${\mathbb {CP}}^k$ these correspond to constant magnetic fields (the field strength components are proportional to the Riemannian curvature which is constant, proportional to the $U(k)$ structure constants, in the tangent frame basis).  In particular, we can choose uniform $U(1)$ or $U(k)$ background magnetic fields,
\beqar
U(1):&~~~~~~~~ &\bar{F} = d \bar{a}= n ~\Omega, \nonumber \\
SU(k):&~~~~~~~~&\bar{F}^a \sim \bar{R}^a \sim f^{a \alpha \beta} e^{\alpha} e^{\beta} \label{background}
\eeqar
where $\Omega$ is the K\"ahler two-form. The choice of the background fields determines how the Landau wavefunctions $\Psi$ transform under right $U(k)$ transformations. In particular,
\beqar
{\hat R}_{k^2 +2k} ~\Psi^J_{m; \alpha} (g) &=& - {n k\over \sqrt{2 k
(k+1)}}~\Psi^J_{m; \alpha} (g) \label{right1} \\
{\hat R}_a ~\Psi^J_{m; \alpha} (g) &=&
(T_a^{\tilde{J}})_{\alpha \beta} \Psi^J_{m; \beta} (g) 
 \label{right2}
\eeqar
where the index $m=1,\cdots, {\rm dim}J$ labels each state within the $SU(k+1)$ representation $J$ and therefore counts the degeneracy of the Landau level. Equation (\ref{right2}) shows that the wavefunctions $\Psi^J_{m; \alpha}$ transform, under right
rotations, as a  representation $\tilde{J}$ 
of
$SU(k)$ in the case of nonabelian backgrounds. The index
$\alpha=1,\cdots {\rm dim}\tilde{J}$ characterizes the nonabelian charge of the underlying fermion fields. 

In the absence of a confining potential, the Hamiltonian $H$ for the Landau problem is proportional to the covariant Laplacian on 
${\mathbb{CP}}^k$, namely
\beq
H \, \Psi = - {1\over 4 M} (\D_{+i} \D_{-i} + \D_{-i} \D_{+i} ) \, \Psi
\label{hamiltonian}
\eeq
which apart from additive constants can be reduced to the form $\sum_{i} {\hat R}_{+i} {\hat R}_{-i}$. Thus the lowest Landau level wavefunctions satisfy the holomorphicity condition
\beq
\hat{R}_{-i} \, \Psi = 0
\label{holo}
\eeq
The conditions (\ref{right1}), (\ref{right2}) and (\ref{holo}) completely fix the representation $J$ and therefore the degeneracy of the lowest Landau level. 

In the case of an abelian magnetic field one can easily write down the lowest Landau level wavefunctions. They are essentially the coherent states for ${\mathbb {CP}}^k$, written explicitly in terms of the holomorphic coordinates $z_i=x_i+iy_i$ as
\beqar
\Psi_{i_1 i_2 \cdots i_k}&=& \sqrt{N} \left[ {n! \over {i_1! i_2! ...i_k!
(n-s)!}}\right]^{\half} ~ {z_1^{i_1} z_2^{i_2}\cdots z_k^{i_k}\over
(1+\bz \cdot z )^{n \over 2}}~,~~~~~~~~~~~~  \nonumber\\
s &=& i_1 +i_2 + \cdots +i_k ~,~~~0\le i_i \le n~~, ~~~  0 \le s \le n \label{wav}
\eeqar 
They form a symmetric, rank $n$ representation $J$ of $SU(k+1)$ whose dimension is 
\beq
N={\rm dim} J = {{(n+k)!} \over {n! k!}} 
\label{dim}
\eeq
The volume element for ${\mathbb{CP}^k}$ (normalized so that the total volume, $\int d\mu$, is $1$) is 
\beq
d\mu = {k! \over \pi^k} {d^2z_1 \cdots d^2z_k
\over (1+ \bz \cdot z)^{k+1}}
\label{vN40}
\eeq

In the case of a $U(1) \times SU(k)$ nonabelian gauge field,
it is convenient to label the irreducible representation of $SU(k+1)_R$ by $(p+l,
q+l')$ corresponding to the tensor \cite{QH3}
\beq
{\cal T}^{a_1...a_q \g_{1}...\g_{l'}}_{b_1...b_p \d_{1}...\d_{l}} \equiv
{\cal T}^{q,l'}_{p,l} \label{7a}
\eeq
where $p,q$ indicate $U(1)$ indices and $l,l'$ indicate $SU(k)$ indices, namely
$a$'s and $b$'s take the value $(k+1)$ and $\gamma$'s and $\delta$'s take
values $1,\cdots,k$. 

The right hypercharge corresponding to (\ref{right2}) is
\beq
\sqrt{2k(k+1)} R_{k^2+2k} = -k(p-q)+l-l' = -nk
\label{7b}
\eeq
The fact that $n$ has to be integer implies that $(l-l')/k$ is an integer,
thus constraining the possible $SU(k)_R$ representations $\tilde{J}$.

Further, as explained in \cite{QH3}, the lowest Landau level states correspond to $q=0,~l=0$ and are described by the representations ${\cal T}^{l'}_p$, where
$p=n-{l' \over k}$ and $l'=jk,~j=1,2,\cdots$. 

\section{Entanglement Entropy for higher dimensional QHE}

Entanglement entropy has been used to explore properties of quantum states in a variety of condensed matter systems. Typically, a system is divided into two subsystems and the entanglement entropy is calculated in terms of the von Neumann entropy of the reduced density matrix of one of the subsystems. For gapped two-dimensional systems, such as QH systems, the leading order contribution to the entanglement entropy is proportional to the perimeter of the boundary separating the two subsystems, in particular $S= a L + \gamma + {\cal{O}}(1/L)$, where $L$ is the length of the boundary, $a$ is a non-universal coefficient and $\gamma$ is a universal quantity referred to as the topological entanglement entropy \cite{kitaev}. The area-law entropy behavior for the two-dimensional fully-filled integer QHE was studied in different geometries analytically for $\nu=1$ and numerically up to $\nu=5$ in \cite{Sierra}; it was found that  $\gamma = 0$ while the coefficient $a$ depends on the filling fraction $\nu$. A general mathematical theorem on the entropy area-law for $\nu=1$ QHE on K\"ahler manifolds is shown in \cite{math}.


In order to calculate the entanglement entropy we divide the system into two regions, $D$ and its complementary $D^C$, and define the reduced density matrix 
\beq
\rho_D = \Tr_{D^C} \vert {GS} \ra \la {GS}\vert
\label{density}
\eeq
where $\vert{GS}\ra$ is the many-body ground state of the system, $\vert{GS} \ra= \prod_{m} c^{\dagger}_m \vert{0}\ra$. 

The entanglement entropy is defined as 
\beq
S= - \Tr \rho_D \log \rho_D
\label{entropy}
\eeq

We choose $D$ to be the spherically symmetric region of ${\mathbb {CP}}^k$ satisfying $z\cdot \bar{z} \le R^2$. For ${\mathbb {CP}}^1 \sim S^2$, this region is a polar cap centered around the north pole and bounded by a latitude angle $\theta$, with $R = \tan \theta/2$ via stereographic projection.

The LLL fermion operator can be expanded as
\beq
\psi = \sum_m c_m \, \Psi_m(z) 
\eeq
where $\Psi_m$ is the single particle wavefunction. We now define the operators $a,~b$ such that
\beqar
a_m = {1\over \sqrt{\lambda_m}} \int_D ~\!\!d\mu ~ \Psi^*_m \psi , &\hskip .2in&
b_m = {1\over \sqrt{1-\lambda_m}} \int_{D^C} ~\!\!d\mu ~ \Psi^*_m \psi \nonumber\\
\lambda_m &=& \int_D ~\Psi^*_m \Psi_m
\nonumber
\eeqar
It is straightforward to check that the pairs $\{ a_m, a^\dagger_m\}$, $\{ b_m, b^\dagger_m\}$ form two independent
fermionic algebras and further
\beq
c_m = \sqrt{\lambda_m}~\, a_m + \sqrt{1-\lambda_m} ~\, b_m \hskip .3in
c^\dagger_m = \sqrt{\lambda_m}~\, a^\dagger_m + \sqrt{1-\lambda_m} ~\, b^\dagger_m 
\label{ab}
\eeq
The operators $a^{\dagger}, b^{\dagger}$ essentially generate the part of the wavefunction which is localized inside $D$ and $D^{C}$ correspondingly. Tracing over $b$'s one finds that the reduced matrix $\rho_D$ takes the form of a $2^N \times 2^N$ block diagonal matrix, in particular 
\beq
\rho_D = \otimes_{m} {\rm diag} (\lambda_m,~1-\lambda_m)
\eeq
The entanglement entropy is then given by
\beq
S = - \Tr \rho_D \log \rho_D=- \sum_{m=1}^{N} \left[ \lambda_m \log \lambda_m + (1 - \lambda_m) \log (1-\lambda_m)\right]
\eeq
$\lambda$'s can also be thought as the eigenvalues of the two-point correlator, with $z,z' \in D$ \cite{Peschel},
\beq
C(r, r') = \sum_{m=1}^{N} \Psi_m^{*} (z) ~\Psi_m (z')
\eeq 

We now proceed to calculate the eigenvalues $\lambda$ and the entanglement entropy for the case of an abelian and nonabelian magnetic field backgrounds. 

\subsection{ $\nu=1$ QHE on ${\mathbb {CP}}^k$ and abelian magnetic field} 

The lowest Landau level wavefunctions for ${\mathbb {CP}}^k$ in the case of an abelian background magnetic field are given in (\ref{wav}). The corresponding eigenvalues $\lambda_s= \int_D ~\Psi^*_s \Psi_s$ are given by
 \beq
 \lambda_{s} = {{ (n+k)!} \over {(s+k-1)! (n-s)!}}
 \int_0^{t_0} dt \, t^{s+k-1} (1-t)^{n-s} 
 \label{beta}
\eeq
where $t_0 = R^2/(1+R^2)$.  
For large $n$, this is amenable to a semiclassical calculation for all $k \ll n$ \cite{VPN}.
For each value of $s= i_1 +i_2 + \cdots +i_k$, the eigenvalue $\lambda_{s}$ has a degeneracy $d_s = {(s+k-1)! / {s! (k-1)!}}$. 

The expression for the entanglement entropy is then 
\beqar
S &= & \sum_{s=0}^{n} ~{(s+k-1)! \over {s! (k-1)!}} ~~ H_s \label{entropy1}\\
H_s &=&  -\lambda_s \log \lambda_s - (1 - \lambda_s) \log (1-\lambda_s) \label{H}
\eeqar
For large $n$ and for any spherical domain $D$ of radius $R$ there is a sharp transition in the value of $\lambda$ around $s^*\sim n t_0$, where $\lambda_{s^*} = 1/2$. Smaller values of $s$ correspond to $\lambda=1$ ($\Psi_s$ are localized within $D$) while bigger values of $s$ correspond to $\lambda =0$ ($\Psi_s$ are localized outside $D$). This pattern is the same for all $k$, when $k \ll n$. As a result only wavefunctions localized very near the boundary of $D$ contribute to the entropy, and $H$ defined in (\ref{H}) is a narrowly peaked Gaussian distribution around $\lambda =1/2$. Figures 1 and 2 clearly illustrate these points. 
\begin{figure}[h]
\begin{center}
\begin{minipage}{5cm}
\scalebox{.5}{\includegraphics{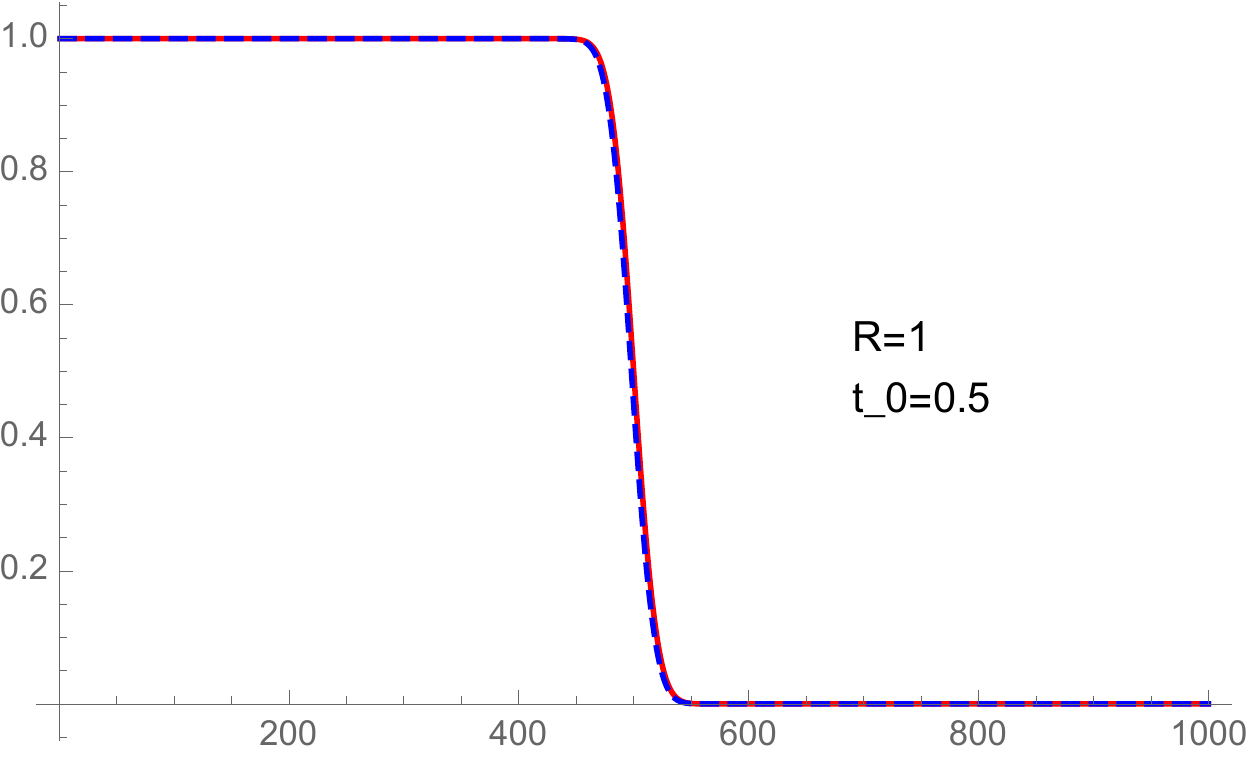}}
\caption{Plots of $\lambda_s$ as a function of $s$ for $k=1$ (red), $k=5$ (blue dashed) and $n = 1000$ and $R=1$}
\label{g1}
\end{minipage}
\hskip 1in
\begin{minipage}{5cm}
\scalebox{.5}{\includegraphics{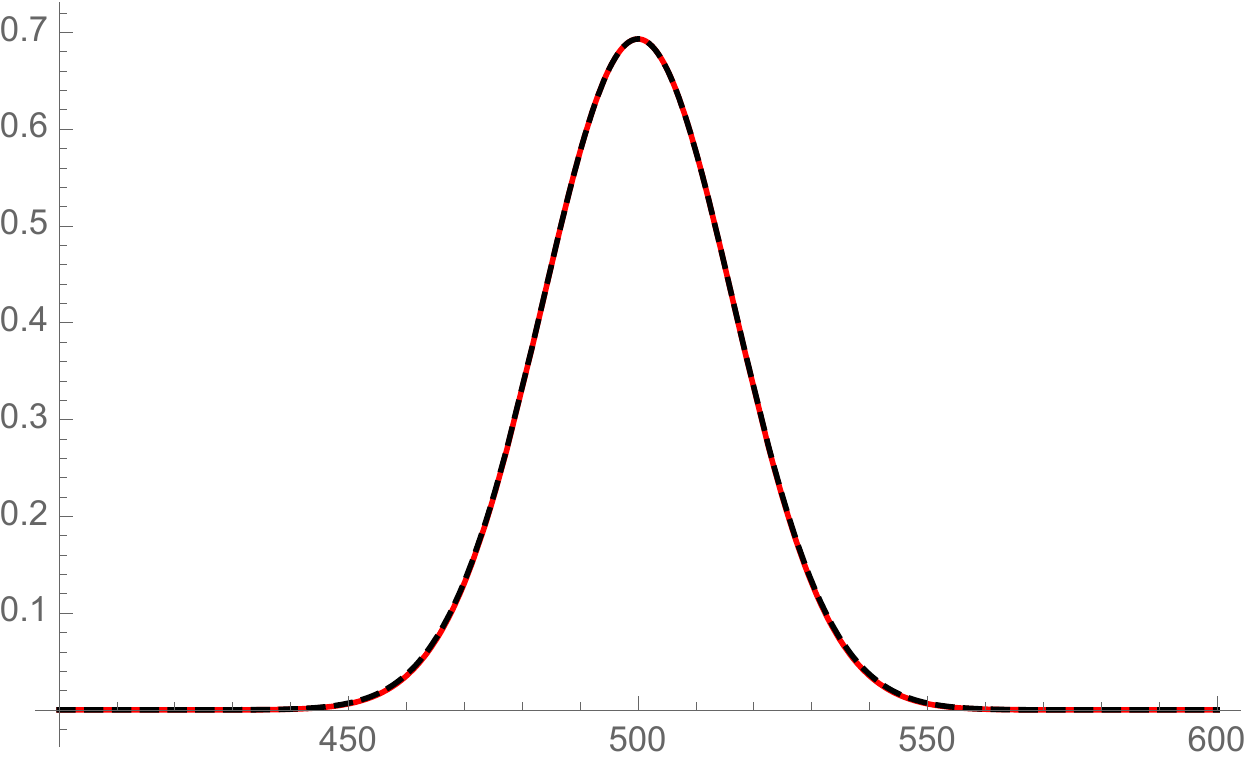} }
\caption{Plots of $H_s$ exact (red) and Gaussian approximation (blue dashed) as a function of $s$ for $k=1$ and $n = 1000$ and $R=1$}
\label{g2}
\end{minipage}
\end{center}
\end{figure}
Extending the semiclassical analysis to the entropy we find that 
 \beq
S  \sim  n^{k -\half} ~{\pi ~( \log 2)^{3/2}  \over {2~k!}} ~2k {R^{2k-1} \over {(1+R^2)^k}} ~\sim ~ c_k ~{\rm Area}
\label{k entropy}
\eeq
where $2k {R^{2k-1} \over {(1+R^2)^k}}$ is the geometric area of the boundary of spherical cap $D$. This agrees with the $k=1$ result in \cite{Sierra}. 

The entropy in (\ref{k entropy}) is proportional to the area of the entangling surface as expected, but the overall coefficient $c_k$ depends on $k$. This dependence goes away though if one expresses the entropy in terms of a "phase-space" area instead of a geometric one. The phase-space volume is proportional to the number of states, namely $V_{\rm phase-space} = { n^k \over k!} ~\int \Omega^k~=~{ n^k \over k!}~\int d \mu$. This then defines a phase-space area for the entangling surface as
\beq
A_{\rm phase-space} = {n^{k-\half}  \over k!} A_{\rm geom} = {n^{k - \half}   \over k!} ~2k{R^{2k-1} \over {(1+R^2)^k}} 
\eeq
Expressing the entanglement entropy (\ref{k entropy}) in terms of this phase-space area we obtain a universal expression valid for any $k$,
\beq
S \sim {\pi \over 2} ( \log 2)^{3/2} ~A_{\rm phase-space}
\eeq

\subsection{ $\nu=1$ QHE on ${\mathbb {CP}}^2$ and nonabelian magnetic field} 

The derivation of the entanglement entropy in the case of a nonabelian background magnetic field is more involved. As mentioned earlier the LLL states form irreducible representations of $SU(k+1)$ of the form ${\cal T}^{l'}_p$, where
$p=n-{l' \over k}$ and $l'=jk,~j=1,2,\cdots$. We will illustrate the calculation of the entanglement entropy for the simplest case of ${\mathbb {CP}}^2$ with a nonabelian magnetic field $U(1) \times SU(2)$ for the lowest value of $l'$ , namely $l'=k=2$ and $p=n-1$. The derivation for other values of $k$ and $l'$ follows similar ideas. The degeneracy of the corresponding LLL is \cite{QH3}
\beq
N = {3 n (n+3) \over 2} 
\label{26}
\eeq
There are now three distinct types of wavefunctions 
\cite{karabali}. Here I will only write down the corresponding expressions for $\lambda$'s for each of the three $SU(2)$ multiplets,
\beqar
\lambda^{(1)}_{s, k=2} &=& \lambda^{(\rm Ab)}_{s+1, k=3} \nonumber \\
\lambda^{(2)}_{s, k=2} &=& { n+3 \over n+1} \lambda^{(\rm Ab)}_{s+1, k=2} -{2 \over n+1}  \lambda^{(\rm Ab)}_{s+1, k=3}  \\
\lambda^{(3)}_{s, k=2} &=& { n+3 \over n+1} \lambda^{(\rm Ab)}_{s+1, k=1} -{2 (n+3) \over {(n+1)(n+2)}}  \lambda^{(\rm Ab)}_{s+1, k=2}  +{2 \over{(n+1)(n+2)}} \lambda^{(\rm Ab)}_{s+1, k=3} \nonumber
\label{62b}
\eeqar
where $\lambda_{s,k}^{(\rm Ab)}$ are the eigenvalues (\ref{beta}) we found for the abelian magnetic field. Each $\lambda^{(i)}$ in (\ref{62b}) has a corresponding degeneracy $s+i$, 1=1,2,3. 

The expression for the entanglement entropy for the nonabelian lowest Landau level states for ${\mathbb {CP}}^2$ can now be written as 
\beqar
S &=&  \sum_{s=0}^{p} \left[ (s+1) H_{s}^{(1)} + (s+2) H_{s}^{(2)} + (s+3) H_{s}^{(3)} \right]  \nonumber \\
&\xrightarrow{large~n} & \sum_{s=0}^{p} \left[ (s+1) H_{s+1, k=3}^{(\rm Ab)} + (s+2) H_{s+1,k=2}^{(\rm Ab)} + (s+3) H_{s+1,k=1}^{(\rm Ab)} \right] \nonumber \\
& \rightarrow & 3~ n^{3/2} ~\pi ~( \log 2)^{3/2}  ~{R^{3} \over {(1+R^2)^2}} = 3 ~S^{(\rm Ab)} 
\eeqar
The overall factor of 3 relating the nonabelian entanglement entropy to the abelian one above has to do with the fact that each lowest Landau state is an $SU(2)$ triplet, ${\rm dim} \tilde{J} = 3$. 
Although the calculation of the entropy in the case of a nonabelian background was explicitly done for $\mathbb{CP}^2$ and the triplet representation, one expects a more general statement to hold. In the large $n$ limit the degeneracy of the LLL in the case of a nonabelian background is \cite{QH3}
\beq
N \sim {\rm dim} \tilde{J} ~{n^k \over k!}
\label{65}
\eeq
The corresponding phase-space volume in this case is $V_{\rm phase-space} = {\rm dim}\tilde{J}~{ n^k \over k!} \int d \mu$ and the corresponding phase-space area is 
\beq
A_{\rm phase-space} = n^{2k-1 \over 2}~  { {\rm dim}\tilde{J} \over k!} A_{\rm geom} = n^{k - \half} { {2 ~{\rm dim}\tilde{J}}   \over (k-1)!} ~{R^{2k-1} \over {(1+R^2)^k}} 
\label{66}
\eeq
Expressed in terms of the phase-space area the overall coefficient in the expression for the entanglement entropy is the same for {\it any abelian or nonabelian background} at large $n$
\beq
S \sim {\pi \over 2} ( \log 2)^{3/2} ~A_{\rm phase-space}
\label{67}
\eeq

\section{Entanglement entropy for higher Landau levels} 

In this section we will consider the entropy for higher Landau levels focusing in particular on some of the differences in the behavior of the eigenvalues $\lambda$ between the lowest Landau level $q=0$, the first excited Landau level $q=1$ and the case of $\nu=2$ where both levels are filled. We will only consider the $k=1$ case, QHE on the sphere. Similar features apply for higher $k$. 

The wavefunctions for the $q$-th Landau level are of the form
\beq
\Psi^J_{m} (g) = \sqrt{N} \, \la{J, m} \vert g \vert {J, n} \ra
\label{68}
\eeq
where $J= {n/2} + q$ and ${\rm dim} J = n+2q+1$. The state $\vert{J, n} \ra$ is not the lowest weight state of the $J$ representation. It can be generated by the action of $\hat{R}_{+}$ on the LLL state. In particular, the correctly normalized wavefunctions of the $q=1$ Landau level are
\beq
\Psi_s^{q=1}= \sqrt{n+3} \sqrt{{(n+1)!} \over {s!(n+2-s)!}}  \Bigl[ {{-(n+2) \bar{z} z^s} \over {\sqrt{(1+z\bar{z})^{n+2}}}} + {s z^{s-1} \over \sqrt{(1+z\bar{z})^{n}} }\Bigr]
\label{72}
\eeq
and the corresponding eigenvalues of the two-point correlator are of the form
\beq
 \lambda_s^{(q=1)} = {{ (n+3)! (n+2)} \over {s! (n+2-s)!}}
 \int_0^{t_0} dt \, t^{s-1} (1-t)^{n-s+1} ~[t-{s \over {n+2}}]^2
 \label{74}
 \eeq
 The eigenvalue $\lambda_s^{(q=1)}$ as a function of $s$ is similar to $\lambda_s^{(q=0)}$ away from the transition region, but it displays a distinct step-like pattern around the value $s^* =t_0~(n+2)$, as shown in Figure 3. The reason for this has to do with the fact that the wavefunctions (\ref{72}) have a node. Since they are generated by the action of $R_{+}$ on the LLL wavefunctions of monopole charge $n+2$ they are necessarily orthogonal to them. Since the LLL wavefunctions are nonzero and have no node, orthogonality requires that the first level Landau wavefunctions must have a node. Higher Landau level wavefunctions acquire more nodes and one expects more steps around the transition region for the corresponding eigenvalues $\lambda$. In fact the $q$-th Landau level states will have $q$ nodes and the profile of the corresponding $\lambda$ will display $q$ distinct steps. 
 
\begin{figure}[h]
\begin{center}
\begin{minipage}{5cm}
\scalebox{.5}{\includegraphics{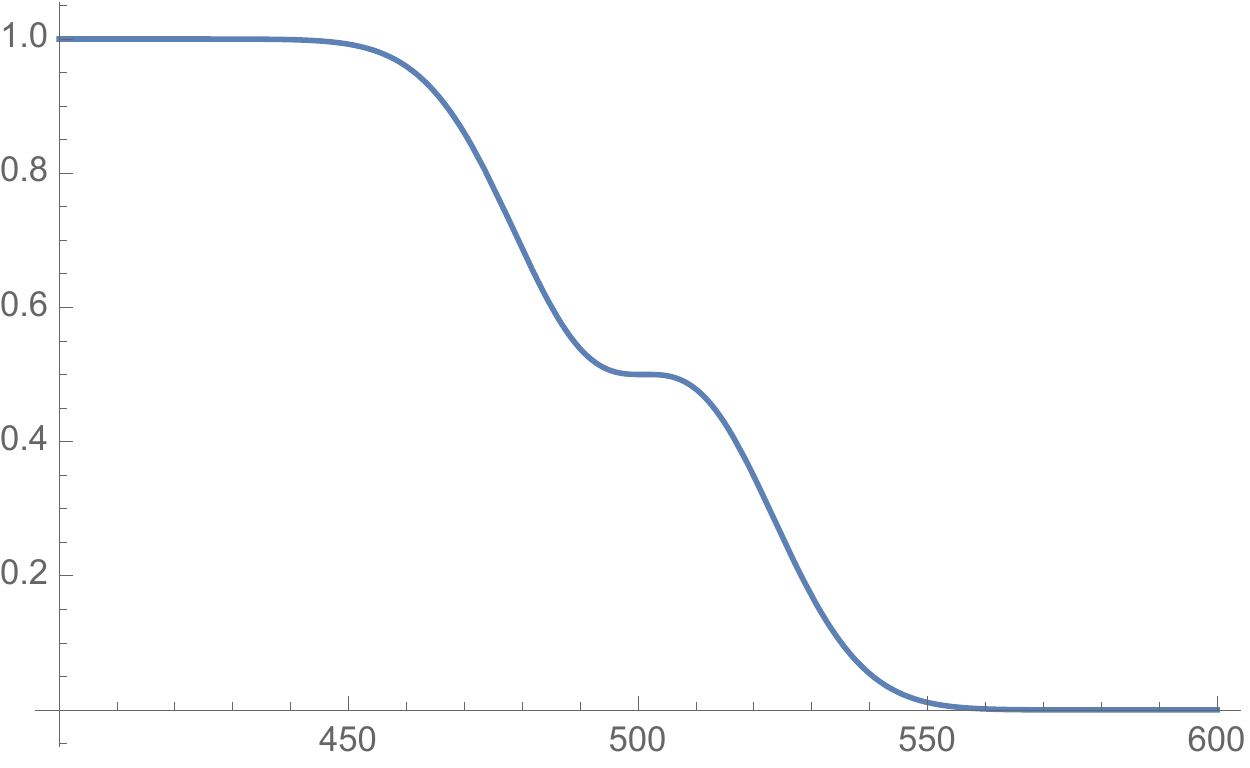}}
\caption{Plot of $\lambda_s^{(q=1)}$ as a function of $s$ for $k=1$, $n = 1000$ and $R=1$}
\label{g1}
\end{minipage}
\hskip 1in
\begin{minipage}{5cm}
\scalebox{.5}{\includegraphics{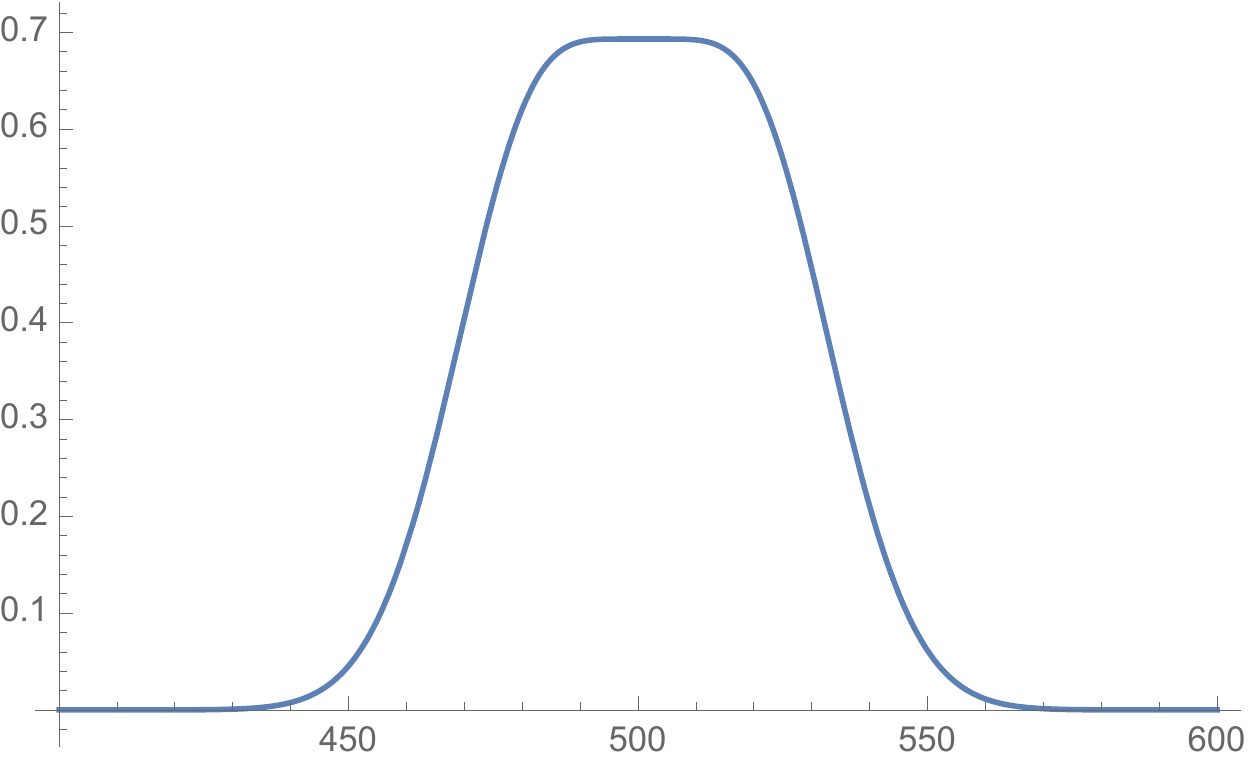} }
\caption{Plot of $H_s^{(q=1)}$ as a function of $s$ for  $n = 1000$ and $R=1$}
\label{g2}
\end{minipage}
\end{center}
\end{figure}
The step-like plateau of $\lambda$ causes the broadening of the entropy per mode $H_s$ around $\lambda=1/2$ which cannot be approximated with a simple narrow Gaussian distribution anymore. Figure 4 displays the plot of $H_s^{(q=1)}$ around $\lambda=1/2$ based on the numerical evaluation of the exact expressions in (\ref{H}) and (\ref{74}). This clearly shows a deviation from the Gaussian distribution (see also Figure 7). 

As a result, the entropy for the first Landau level is larger than the entropy for the LLL even though the number of states are approximately the same at large $n$ ($n+1$ states for $q=0$ versus $n+3$ states for $q=1$).
A numerical evaluation of the entropy shows that it obeys an area law and it gives
\beq
S^{(q=1)} =1.65 ~S^{(q=0)}
\label{82}
\eeq

When both $q=0$ and $q=1$ levels are filled, namely $\nu=2$, the situation is more involved as there are overlaps between the wavefunctions of different Landau levels. 
The two-point correlator is now given by
\beq
C(r,r') = \sum_{s=0}^{n} \Psi^{*0}_s (r) \Psi^{0}_s (r') + \sum_{s=0}^{n+2} \Psi^{*1}_s (r) \Psi^{1}_s (r')
\label{84}
\eeq
There are $2n+4$ eigenvalues : $\lambda_0^1~, \tilde{\lambda}_{s}^{\pm} ~,~ \lambda^1_{n+2}$, where $s=0,\cdots,n$ and 
\beq
\tilde{\lambda}_s^ {\pm} = {{\lambda_s^0 + \lambda_{s+1}^1 \pm \sqrt{ (\lambda_s^0-\lambda_{s+1}^1)^2 + 4 (\delta\lambda)^2_{s,s+1} }}\over 2}
\label{86}
\eeq
where $\lambda^{0}~,~\lambda^{1}$ are the eigenvalues we derived earlier for the lowest and first Landau level and $\delta\lambda$ is the overlap 
\beq
\delta\lambda_{s,s'} = \int_D \Psi^{*(q=0)}_s (r)~\Psi^{(q=1)}_{s'} (r) d\mu
\label{83}
\eeq
The interesting feature here is that once both Landau levels are included the step like pattern in the profile of $\lambda^1$ disappears. The profile of the new $\tilde{\lambda}^{\pm}$ resembles that of $\lambda^0$ but shifted with respect to $\lambda^0$, see Figure 5. As a result, the corresponding entropies per mode $\tilde{H}^{\pm}_s$, where
\beq
\tilde{H}^{\pm}_s  = -\tilde{\lambda}^{\pm}_s \log \tilde{\lambda}^{\pm}_s - (1 - \tilde{\lambda}^{\pm}_s) \log (1-\tilde{\lambda}^{\pm}_s) \label{87}
\eeq
are Gaussian distributions centered around the value of s for which $\tilde{\lambda}^{\pm} =1/2$ as shown in Figure 6.
\begin{figure}[t]
\begin{center}
\begin{minipage}{5cm}
\scalebox{.5}{\includegraphics{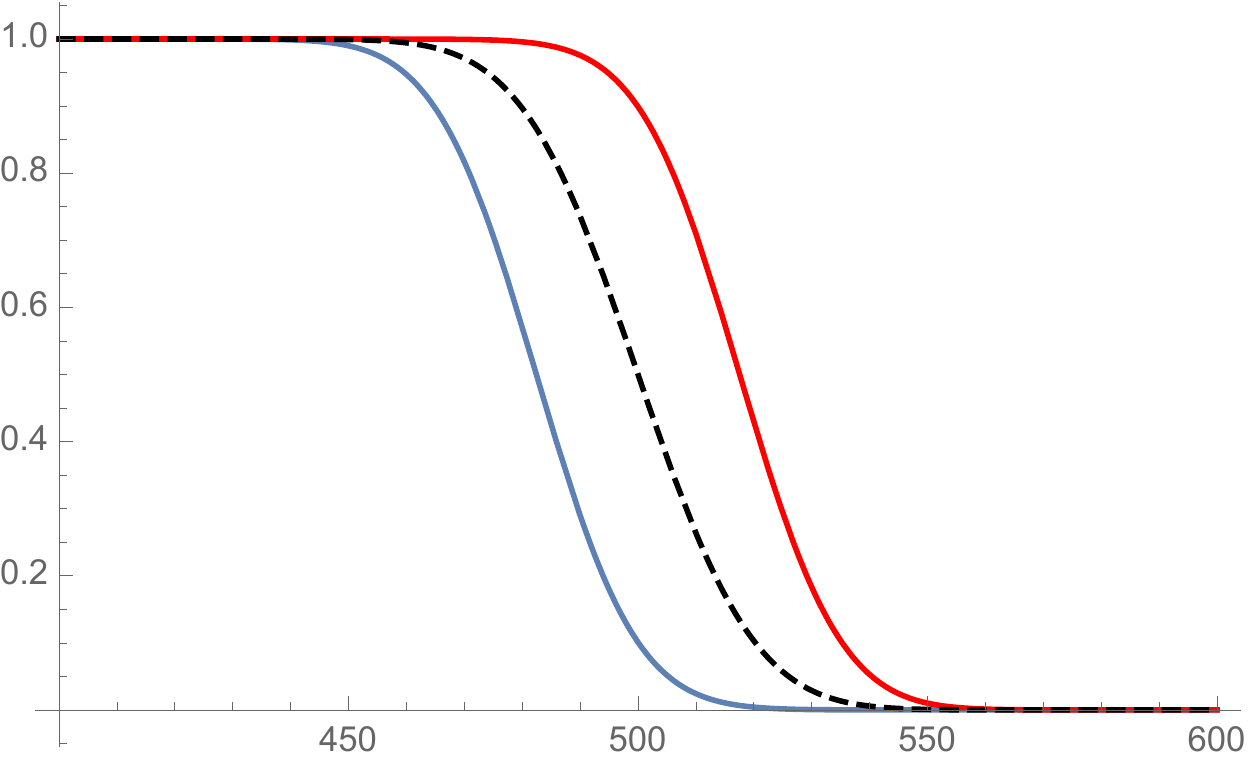}}
\caption{Plots of $\tilde{\lambda}^+$ (red to the right), $\tilde{\lambda}^-$ (blue to the left) as functions of $s$ for  $n = 1000$ and $R=1$, compared to $\lambda^0$ (dashed, center)}
\label{g1}
\end{minipage}
\hskip 1in
\begin{minipage}{5cm}
\scalebox{.5}{\includegraphics{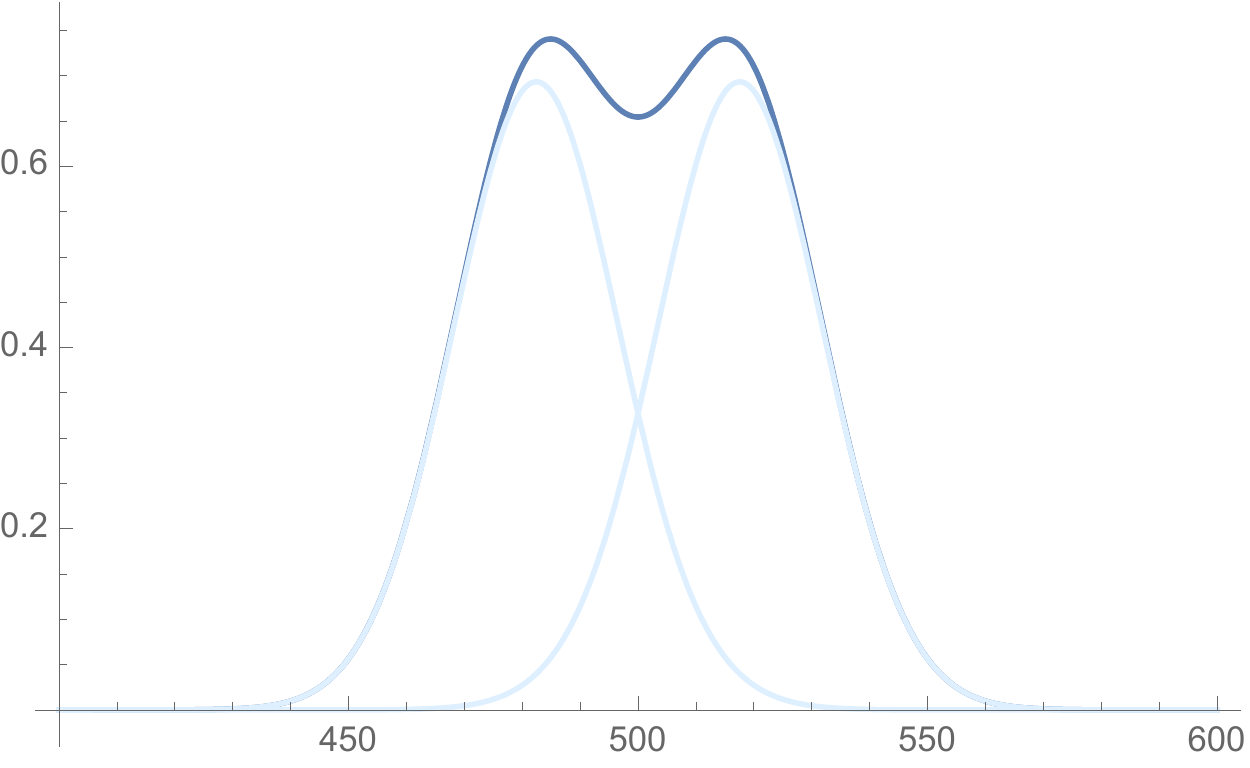} }
\caption{Plot of $\tilde{H}^+ + \tilde{H}^-$  as function of $s$ for  $n = 1000$ and $R=1$}
\label{g2}
\end{minipage}
\end{center}
\end{figure}
\begin{figure}[h]
\begin{center}
\scalebox{.5}{\includegraphics{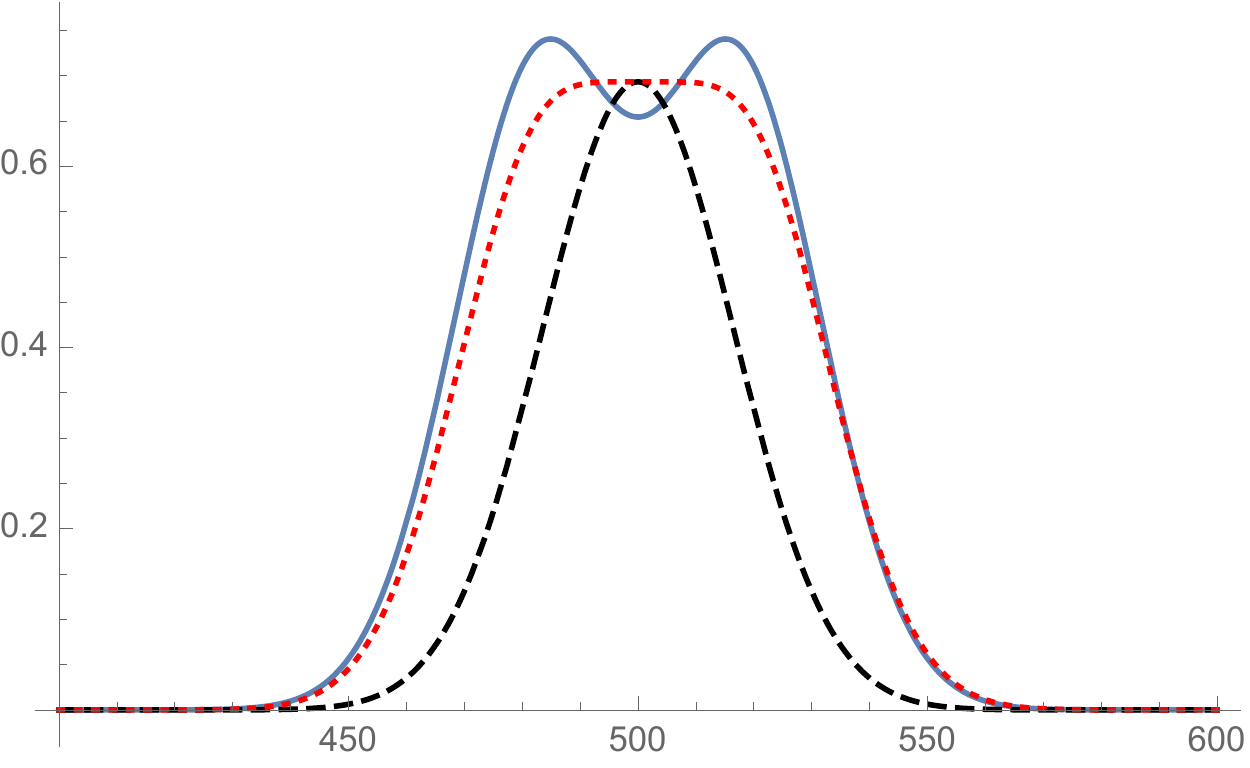}}
\caption{Plots of $H^{(\nu=1)}$ (black, dashed), $H^{(q=1)}$ (red, dotted) and $H^{(\nu=2)}$ (blue, solid) as functions of $s$ for  $n = 1000$ and $R=1$}
\label{g1}
\end{center}
\end{figure}

Figure 7 shows a comparison between $H^{(q=0)}~, H^{(q=1)}$ and $H^{(\nu=2)}$ which explains the differences in the values of the corresponding entropies, namely
\beq
S^{(\nu=2)} > S^{(q=1)} > S^{(\nu=1)}
\label{88}
\eeq
A numerical evaluation of the entropy for the $\nu=2$ case gives
\beq
S^{(\nu=2)} = 1.76~S^{(\nu=1)}
\label{89}
\eeq
This agrees with the result in \cite{Sierra}.

\section{Summary, Comments}

I have presented various aspects of higher dimensional integer quantum Hall effect defined on ${\mathbb{CP}}^k$ with a particular emphasis on the issue of entanglement entropy. The connections to noncommutativity, bosonization and effective actions of such systems are quite general and apply for any K\"ahler manifold. In fact the bosonization approach outlined in section 2 and the corresponding edge effective actions are also relevant for bosonization of fermionic systems around a fermi surface (phase-space droplets) \cite{poly1}. 

Regarding entanglement entropy, we found, based on a semiclassical analysis, that for $\nu =1$ the entropy is proportional to the area of the entangling surface as in the lower dimensional case. Furthermore, we found an interesting universality relation, namely, if the area law for the entropy is expressed in terms of the phase-space area of the entangling surface, the overall proportionality constant remains the same for any dimension and for any background, abelian or nonabelian. The semiclassical analytical calculation breaks down for higher Landau levels.

In the presence of edge degrees of freedom the entanglement entropy for two-dimensional systems develops subleading logarithmic contributions $S_{{\rm edge}} \sim {c \over 6} \log l$, where $c$ is the central charge of the gapless edge modes and $l$ is the segment of the entangling boundary intersecting the chiral droplet. This was indeed confirmed in the case of the two-dimensional $\nu=1$ QHE droplets whose edge dynamics is characterized by a chiral scalar of central charge $c=1$  \cite{estienne} and was recently extended to the case of a four-dimensional QHE analog with abelian magnetic field \cite{estienne2}. We have previously analyzed in detail higher dimensional QHE droplets, their corresponding edge effective actions and their spectrum \cite{QH2,QH3}. In the case of the abelian chiral droplets the effective action is essentially that of a collection of one-dimensional chiral scalars, whose direction of propagation along the surface of the droplet is uniquely determined by the gradient of the confining potential and the K\"ahler form of the manifold \cite{QH2}. Because of that, the theory, although higher dimensional, is similar to a collection of (1+1) dimensional conformal field theories, and therefore one expects the edge entanglement entropy to exhibit a logarithmic behavior and a dependence on the central charge similar to the case in two dimensions. The analysis of the same question for nonabelian chiral droplets is more challenging and interesting, since the conformal properties of the nonabelian edge action are not obvious.

Another interesting question is how the entanglement entropy changes in the presence of gauge and gravitational fluctuations \cite{VPN}. This is the subject of the next talk by Nair \cite{nairtalk}. 

{\it Acknowledgements}: I thank the organizers of the CORFU21 workshop for the invitation and the hospitality. This research was supported by the NSF grant PHY-1915053 
and PSC-CUNY awards.


\begin{thebibliography}{99}

\bibitem{girvin} R.E. Prange and S.M. Girvin, {\it The Quantum Hall Effect}, 2nd ed. (Springer,
Berlin, 2012); S. Girvin, "The quantum Hall effect: novel
excitations and broken symmetries", cond-mat/9907002 and references
therein. 
\bibitem{HZ} S.C. Zhang and J.P. Hu, "A four dimensional generalization of the quantum Hall effect", {\it Science} {\bf 294} (2001) 823; J.P. Hu and S.C. Zhang, "Collective excitations at the boundary of 4D quantum Hall droplet", {\it Phys. Rev.} {\bf B66} (2002) 125301.
\bibitem{QH1} D. Karabali and V.P. Nair, "Quantum Hall effect in higher dimensions", {\it Nucl. Phys.} {\bf B641} (2002) 533.
\bibitem{QH2} D. Karabali and V.P. Nair, "The effective action for edge states in higher dimensional quantum Hall systems", {\it Nucl. Phys.} {\bf B679} (2004) 427.
\bibitem{QH3} D. Karabali and V.P. Nair, "Edge states for quantum Hall droplets in higher dimensions and a generalized WZW model", {\it Nucl. Phys.} {\bf B697} (2004) 513.
\bibitem{QH4} D. Karabali, V.P. Nair and S. Randjbar-Daemi, "Fuzzy spaces, the M(atrix) model and the quantum Hall effect", hep-th/0407007; D. Karabali and V.P. Nair, "Quantum Hall Effect in Higher Dimensions, Matrix Models and Fuzzy Geometry", {\it J.Phys.} {\bf  A39} (2006) 12735.
\bibitem{QH5} D. Karabali, "Electromagnetic interactions of higher dimensional quantum Hall droplets", {\it Nucl. Phys. }{\bf B726} (2005) 407.
\bibitem{QH6} D. Karabali, "Bosonization of the lowest Landau level in arbitrary dimensions: edge and bulk dynamics", {\it Nucl. Phys.} {\bf B750} (2006) 265; "Quantum Hall Effect, Bosonization and Chiral Actions in Higher Dimensions" hep-th/1005.4341.
\bibitem{QHEaction}  D. Karabali and V.P. Nair,  "The geometry of quantum Hall effect: An action for all dimensions", {\it Phys. Rev.}
{\bf D94} (2016) 024022; "Role of the spin connection in quantum Hall effect:
A Perspective from geometric quantization",
{\it Phys. Rev.} {\bf D94} (2016) 064057.
\bibitem{karabali} D. Karabali, "Entanglement entropy for integer quantum Hall effect in two and higher dimensions", {\it Phys. Rev.} {\bf D102} (2020) 025016.


\bibitem{Haldane} F. D. M. Haldane, "Fractional Quantization of the Hall Effect: A Hierarchy of Incompressible Quantum Fluid States", {\it Phys. Rev. Lett.} {\bf  51} (1983) 605.

\bibitem{others}H. Elvang, J. Polchinski, "The quantum Hall effect on $R^4$", {\it C.R. Physics} {\bf 4}(2003) 405; B.A. Bernevig et al.,
"Effective field theory description of the higher dimensional quantum Hall liquid", 
{ \it Ann. Phys.} {\bf 300} (2002) 185;
B. A. Bernevig, J.P. Hu, N. Toumbas, S.C. Zhang, "Eight-dimensional quantum Hall effect and octonions", { \it Phys. Rev. Lett.}
{\bf 91} (2003) 236803; 
 G. Meng, "Geometric construction of the quantum Hall effect in all even dimensions", { \it J. Phys.} {\bf A36}  (2003) 9415;  V.P. Nair and S. Randjbar-Daemi, "Quantum Hall effect on $S^3$, edge states and fuzzy $S^3 / Z_2$", {\it Nucl. Phys.}
{\bf B679} (2004) 447 ; A. Jellal, "Quantum Hall effect on higher-dimensional spaces", {\it Nucl. Phys.} {\bf B725} (2005) 554;
K. Hasebe, "Higher dimensional quantum Hall effect as A-class topological insulator", { \it Nucl. Phys.} {\bf B886} (2014) 952; "Higher (Odd) Dimensional Quantum Hall Effect and Extended Dimensional Hierarchy", {\it Nucl. Phys.} {\bf B920} (2017) 475; U. H. Coskun, S. Kurkcuoglu, G. C. Toga, "Quantum Hall Effect on Odd Spheres", {\it Phys. Rev.} {\bf D95} (2017) 065021.


\bibitem{exp} Y. E. Kraus, Z. Ringel, and O. Zilberberg, "Four-Dimensional Quantum Hall Effect in a Two-Dimensional Quasicrystal”, {\it Phys. Rev. Lett.} {\bf 111} (2013) 226401; H. M. Price et al.,
"Four-Dimensional Quantum Hall Effect with Ultracold Atoms”, {\it Phys. Rev. Lett.} {\bf 115} (2015) 195303; T. Ozawa et al.,
"Synthetic
dimensions in integrated photonics: From optical isolation to four-dimensional quantum Hall physics”, {\it Phys. Rev.} {\bf A93} (2016) 043827; O. Zilberberg et al.,
"Photonic topological boundary pumping as a probe of 4D quantum Hall physics”, {\it Nature} {\bf 553} (2018) 59; M. Lohse et al.,
"Exploring 4D
quantum Hall physics with a 2D topological charge pump,” {\it Nature} {\bf 553} (2018) 55.

\bibitem{sakita2} B. Sakita, "$W_{\infty}$ Gauge Transformations and the Electromagnetic Interactions of Electrons in the Lowest Landau Level", {\it Phys. Lett.} {\bf B315} (1993) 124; "Collective variables of fermions and bosonization", {\it Phys. Lett.} {\bf B387} (1996) 118.

\bibitem{otherQHE1} A.G. Abanov and A. Gromov, "Electromagnetic and gravitational responses of two-dimensional non-interacting electrons in background magnetic field", {\it Phys. Rev.}  { \bf B90} (2014) 014435; A. Gromov and A. Abanov, "Density curvature response and gravitational anomaly", {\it Phys. Rev. Lett.} { \bf 113} (2014) 266802 .

\bibitem{otherQHE2} T. Can, M. Laskin and P. Wiegmann, "Fractional quantum Hall effect in a curved space: Gravitational anomaly and electromagnetic response", {\it Phys. Rev. Lett.} { \bf 113} (2014) 046803; "Geometry of quantum Hall states: Gravitational anomaly and kinetic coefficients", {\it Ann. Phys.} { \bf 362} (2015) 752.









\bibitem{kitaev} A. Kitaev and J. Preskill, "Topological entanglement entropy", {\it \PRL}~{\bf 96} (2006) 110404; M. Levin and X.G. Wen, "Detecting topological order in a ground state wave function", {\it \PRL}~{\bf 96} (2006) 110405. 

\bibitem{Sierra} I.D. Rodriguez and G. Sierra, "Entanglement entropy of integer quantum Hall states", {\it \PR}~{\bf B80} (2009) 153303; "Entanglement entropy of integer quantum Hall states in polygonal domains", {\it J. Stat. Mech.} {\bf 12} (2010) 12033. 

\bibitem{math} L. Charles and B. Estienne, "Entanglement entropy and Berezin-Toeplitz operators", {\it \CMP}~{\bf 376} (2020) 521.

\bibitem{Peschel} I. Peschel, "Calculation of reduced density matrices from correlation functions", {\it J. Phys. A: Math. Gen.} {\bf 36} (2003) L205 .


\bibitem{VPN} V.P. Nair, "Entanglement for Quantum Hall states and a Generalized Chern-Simons Form", {\it \PR}~{\bf D101} (2020) 125021.

\bibitem{poly1} A.P. Polychronakos, "Chiral actions from phase space (quantum Hall) droplets", {\it Nucl. Phys.} {\bf B705} (2005) 457 ; "Kac-Moody theories for colored phase space (quantum Hall) droplets", {\it Nucl. Phys.} {\bf B711} (2005) 505.


\bibitem{estienne} B. Estienne and J-M. Stephan, "Entanglement spectroscopy of chiral edge modes in the quantum Hall effect", {\it \PR}~{\bf B101} (2020) 115136; P-G. Rozon, P-A. Bolteau and W. Witczak-Krempa, "Geometric entanglement in integer quantum Hall states with boundaries", {\it Phys. Rev.} {\bf B102} (2020) 155417.

\bibitem{estienne2} B. Estienne, B. Oblak and J-M. Stephan, "Ergodic edge modes in the 4D quantum Hall effect", {\it SciPost Phys.} {\bf 11} (2021) 016.

\bibitem{nairtalk} V.P. Nair, "Entanglement Entropy and Matter-Gravity Couplings for
Fuzzy Geometry", Corfu2021 Proceedings, arXiv:2203.14351.














\end{thebibliography}
\end{document}